\documentclass[prb]{revtex4}
\usepackage{amssymb}
\usepackage{graphicx}
\usepackage{array}
\usepackage{hhline}
\usepackage{longtable}

\begin{document}
\title{Nonequilibrium dynamics of a mixed spin-1/2 and spin-3/2 Ising ferrimagnetic system with a time
dependent oscillating magnetic field source}
\author{Erol Vatansever}
\affiliation{Dokuz Eyl\"{u}l University, Graduate School of Natural and Applied Sciences, TR-35160 Izmir, Turkey}
\author{Hamza Polat}
\email{hamza.polat@deu.edu.tr}
\affiliation{Department of Physics, Dokuz Eyl\"{u}l University, TR-35160 Izmir, Turkey}
\date{\today}
\begin{abstract}
Nonequilibrium phase transition properties of a mixed Ising ferrimagnetic
model consisting of spin-1/2 and spin-3/2 on a square lattice under the
existence of a time dependent oscillating magnetic field have been
investigated by making use of Monte Carlo simulations with single-spin flip
Metropolis algorithm. A complete picture of dynamic phase boundary and magnetization
profiles have been illustrated and the conditions of a dynamic compensation
behavior have been discussed in detail. According to our simulation results,
the considered system does not point out a dynamic compensation behavior, when it only
includes the nearest-neighbor interaction, single-ion anisotropy
and an oscillating magnetic field source. As the next-nearest-neighbor
interaction between the spins-1/2 takes into account and exceeds a characteristic value which
sensitively depends upon values of single-ion anisotropy and only of
amplitude of external magnetic field, a dynamic compensation behavior occurs
in the system. Finally, it is reported that it has not been found any
evidence of dynamically first-order phase transition between
dynamically ordered and disordered phases,  which conflicts with the recently
published molecular field investigation, for a wide range of selected system
parameters.

\end{abstract}
\pacs{05.50.+q, 05.70.Fh, 64.60.Ht, 75.50.Gg}
\keywords{Lattice Theory, Phase Transitions in Statistical Mechanics and Thermodynamics,
Dynamic critical behavior, Ferrimagnetics} 
\maketitle
\section{Introduction}
The phenomenon of ferrimagnetism is related to the counteraction of opposite magnetic
moments with unequal magnitudes located on different sublattices. Ferrimagnetic
materials have, under certain conditions, a compensation temperature at which the
resultant magnetization vanishes below its critical temperature \cite{Neel}.
Recently, it has been both experimentally and theoretically
shown that the coercive field exhibits a rapid increase at the
compensation point \cite{Hansen, Buendia1}. It is obvious that such kind of point has a technological
importance \cite{Mansuripur1, Shieh}, because at this  point only a
small  driving field is required to change the sign of the resultant magnetization. Due to the recent
developments in experimental techniques, scientists begin to synthesize new classes of
molecular-based magnets \cite{Manriquez, Morin, Du}. For instance, it has been shown
that the saturation magnetization, chemical analysis and infrared spectrum analysis
of $\mathrm{V(TCNE)_{x}.y(solvent)}$, where $\mathrm{TCNE}$ is tetracyanoethylene, are
consistent with a ferrimagnet with spin-3/2 at the vanadium site and a
spin of 1/2 at the TCNE sites with $\mathrm{x\sim 2}$  \cite{Morin}. In this regard, it is possible
to mention that the theoretical models referring the mixed systems are of great importance
since they  are well adopted to study and to provide deeper
understanding of certain type of ferrimagnetism \cite{Neel}.

From the theoretical point of view, a great deal number of studies
have been realized to get  a clear idea about the magnetic properties
of mixed spin-1/2 and spin-3/2  ferrimagnetic Ising systems.
In order to have a general overview about it, it is beneficial to classify the studies in two
categories based on the investigation of equilibrium and nonequilibrium
phase transition properties  of such type of mixed spin systems. In the former group,
static or equilibrium properties of these type of systems have been analyzed within the several
frameworks such as exact \cite{Domb, Goncalves},  effective field theory
with correlations \cite{Kaneyoshi1, Kaneyoshi2, Benayad, Bobak1, Bobak2, Jiang1,
Liang1, Aouzi, Essaoudi, Liang2, Yigit1, Yigit2}, Bethe lattice \cite{Ekiz1, Ekiz2,
Albayrak1, Albayrak2},  exact star-triangle mapping
transformation \cite{Jascur}, high temperature
series expansion method \cite{Hunter}, multisublattice Green-function
technique \cite{Li}, Oguchi approximation \cite{Bobak3} as well as Monte Carlo
simulation \cite{Buendia2}. It is underlined in the some studies noted above
 that  when the system includes only the nearest-neighbor
interaction between spins and the single-ion anisotropy, the temperature variation of
resultant magnetization does not exhibit a compensation behavior.
In contrary to this, when the next-nearest neighbor interaction
between spins-1/2 takes into account and exceeds a minimum value which
depends upon the other system parameters, the ferrimagnetic system reveals a
compensation treatment which can not be observed in single-spin
Ising systems.

Magnetically interacting system under the influence of a magnetic field varying sinusoidally in time exhibits
two important striking phenomena: Nonequilibrium phase transitions and dynamic hysteresis behavior.
Nowadays, these types of nonequilibrium systems are in the center of scientists
because they have exotic, unusual and interesting behaviors. For example, the universality
classes of the Ising model and its variations under a time dependent driving field are
different from its equilibrium counterparts \cite{Chakrabarti, Acharyya1, Park}. It is
possible to emphasize that  nonequilibrium phase transitions originate due to a competition
between time scales of the relaxation time of the system  and oscillating period of the external applied field.
For the high temperatures and high amplitudes of the periodically varying magnetic field,
the simple kinetic ferromagnetic system exists in dynamically  disordered phase where
the time dependent magnetization oscillates around value of zero and is able to follow the external applied
magnetic field with some delay, whereas it oscillates around a non-zero value
which indicates a dynamically ordered phase for low temperatures and small magnetic
field amplitudes \cite{Tome}. The physical mechanism described briefly above points out the existence of
a dynamic phase transition (DPT) \cite{Chakrabarti, Tome, Lo}.

DPTs  and hysteresis behaviors can also be observed experimentally. For example,
by benefiting from surface magneto-optic Kerr effect (MOKE), dynamic
scaling of magnetic hysteresis in ultrathin ferromagnetic Fe/Au(001) films has been
studied, and it is reported that the dispersion of hysteresis loop area of studied system obeys
to a power law behavior\cite{He}. A comprehensive study, which includes
the hysteresis loop measurement of well-characterized ultrathin Fe films grown on flat
and stepped W(110) surfaces, has been done by using MOKE, and prominent experimental
observations are reported in Ref. \cite{Shin}.  In addition to these pioneering works  mentioned briefly above,
to the best of our knowledge, there exist a number of experimental studies regarding the nonequilibrium
properties  of different types of magnetic materials such as Co films on
a Cu(001) surface \cite{Jiang2}, polycrystalline $\mathrm{Ni_{80}Fe_{20}}$
films \cite{Choi}, epitaxial Fe/GaAs(001) thin films \cite{Lee}, $\mathrm{Fe_{0.42}Zn_{0.58}F_{2}}$ \cite{Rivera},
finemet thin films with composition $\mathrm{Fe_{73.5}Cu_{1}Nb_{3}Si_{13.5}B_{9}}$ \cite{Santi}, $\mathrm{[Co/Pt]_{3}}$
magnetic  multilayers with strong perpendicular anisotropy \cite{Robb} as well as assembly of
paramagnetic colloids \cite{Tierno}. Based upon the detailed experimental investigations, it
has been discovered that experimental nonequilibrium dynamics of considered real magnetic systems strongly resemble the
dynamic behavior predicted from theoretical calculations of a kinetic Ising model. From
this point of view, it is possible to see that there exists an impressive
evidence of qualitative consistency between theoretical and experimental investigations.

On the other hand, in the latter group there exists a limited
number of nonequilibrium studies concerning the influences of time varying magnetic
field on the mixed spin-1/2 and spin-3/2 Ising
ferrimagnetic model. For instance, thermal and magnetic properties of a
mixed Ising ferrimagnetic model consisting of spin-1/2 and spin-3/2 on a
square lattice have been analyzed by making use of Glauber-type
stochastic process \cite{Glauber}. It has been reported that the studied system
always exhibits a dynamic tricritical point in amplitude of external applied field and temperature plane, but it does not show in
the single-ion anisotropy and temperature plane for low values of amplitude of field \cite{Deviren}.
Following the same methodology, a similar study
has been done to shed some light on what happens when an oscillating magnetic
field is applied to the mixed spin-1/2 and spin-3/2 Ising model
on alternate layers of hexagonal lattice. It has been found that depending on
the Hamiltonian parameters, the system  presents dynamic multicritical as well as
compensation behaviors \cite{Keskin}. However, the aforementioned studies are
mainly based on molecular field theory. It is a well known fact that, in molecular field theory, spin
fluctuations are ignored and the obtained results do not have any microscopic information
details of system. From this point view,
in order to obtain the true dynamics of a mixed spin-1/2 and
spin-3/2 Ising ferrimagnetic system on a square lattice under the presence of
a time dependent oscillating magnetic field, we intend to use of Monte Carlo simulation
technique which takes  into account the thermal fluctuations, and in this way,
non-artificial results can be obtained.

The outline of the paper is as follows: In section \ref{formulation} we
briefly present our model. Section \ref{discussion} is dedicated to the results and
discussion, and finally section \ref{conclusions} contains our conclusions.

\section{Formulation}\label{formulation}
We consider a two-dimensional kinetic Ising ferrimagnetic system with mixed spins
of $\sigma=1/2$ and $S=3/2$ defined on a square lattice, and the system is
exposed to a time dependent magnetic field source.  The Hamiltonian describing
our model is given by

\begin{equation}\label{eq1}
\mathcal{H}=-J_{1}\sum_{\langle nn \rangle }\sigma_{i}^{A}S_{j}^{B}-J_{2}\sum_{\langle nnn \rangle}\sigma_{i}^{A}\sigma_{k}^{A}-
D\sum_{j}(S_{j}^{B})^{2}-H(t)\left(\sum_{i}\sigma_{i}^{A}+\sum_{j}S_{j}^{B}\right)
\end{equation}
where the $\sigma_{i}=\pm 1/2$, and $S_{j}=\pm 3/2, \pm 1/2$ are the
Ising spins on the sites of the sublattices A and B, respectively.
First and second sums in Eq. (\ref{eq1}) are over the  nearest- and next-nearest
neighbor pairs of spins, respectively. We assume $J_{1}<0$ such that the
exchange interaction between nearest neighbours  is antiferromagnetic. $J_{2}$ is the
exchange interaction parameter between pairs of next-nearest neighbors of spins
located on sublattice A, and $D$ is single ion-anisotropy term which affects only $S=3/2$ spins
located on sublattice B.  The time varying sinusoidal magnetic field is as following
\begin{equation}\label{eq2}
  H(t)=h_{0}\sin(\omega t)
\end{equation}
here, $h_{0}$ and $\omega$ are amplitude and angular frequency of the external
field, respectively. The period of the oscillating magnetic field is given by $\tau=2\pi/\omega$.

The linear dimension of the lattice is selected as $\mathrm{L=40}$ through  all simulations, and
Monte Carlo simulation based on Metropolis algorithm \cite{Binder}  is applied to the kinetic mixed Ising
ferrimagnetic system on a $40 \times 40$ square lattice with periodic  boundary conditions in all directions.  
Configurations were generated  by selecting the sites sequentially  through the lattice
and making single-spin-flip attempts, which were accepted or rejected according to the Metropolis algorithm.
Data were generated over 50 independent samples realizations by running the simulations for 60000 MC  steps per site after
discarding the first 20000 steps. This amount of transient steps is found to be sufficient for thermalization for
the whole range of the parameter sets. Error bars are found by using Jacknife method \cite{Newman}. Because
the calculated errors are usually smaller than the sizes of the symbols in the obtained figures, they
have not been given in this study.

The instantaneous values of the sublattice magnetizations $M_{A}$ and $M_{B}$, and also the total
magnetization $M_{T}$ at the time t are defined as
\begin{equation}\label{eq3}
M_{A}(t)=\frac{2}{L^2}\sum_{i\in A}\sigma_{i}^{A},\quad \quad \quad M_{B}(t)=\frac{2}{L^2}\sum_{j\in B}S_{j}^{B},
\quad \quad \quad   M_{T}(t)=\frac{M_{A}(t)+M_{B}(t)}{2}.
\end{equation}
By benefiting from the instantaneous magnetizations over a full period of oscillating
magnetic field, we obtain the dynamic order parameters as follows

\begin{equation}\label{eq4}
Q_{A}=\frac{1}{\tau}\oint M_{A}(t)dt, \quad \quad \quad Q_{B}=\frac{1}{\tau}\oint M_{B}(t)dt,
\quad \quad \quad Q_{t}=\frac{1}{\tau}\oint M_{T}(t)dt,
\end{equation}
where $Q_{A}$, $Q_{B}$ and $Q_{t}$ denote the dynamic order parameters corresponding to
the sublattices $A$ and $B$, and the overall lattice, respectively. To determine the dynamic compensation
temperature $T_{comp}$ from the computed magnetization data, the intersection point of the absolute
values of the dynamic sublattice magnetizations was found using

\begin{equation}\label{eq5}
|Q_{A}(T_{comp})|=|Q_{B}(T_{comp})|,
\end{equation}
\begin{equation}\label{eq6}
sign(Q_{A}(T_{comp}))=-sign(Q_{B}(T_{comp})),
\end{equation}
with $T_{comp}<T_{c}$, where $T_{c}$ is the dynamic critical temperature. We also calculate the
time average of the cooperative part of energy of the kinetic mixed Ising ferrimagnetic
system over a full cycle of the  magnetic field as follow \cite{Acharyya2}
\begin{equation}\label{eq7}
E_{coop}=-\frac{1}{L^2 \tau}\oint\left(J_{1}\sum_{\langle nn \rangle }\sigma_{i}^{A}S_{j}^{B}+
J_{2}\sum_{\langle nnn \rangle}\sigma_{i}^{A}\sigma_{k}^{A}+D\sum_{j}(S_{j}^{B})^{2}\right)dt.
\end{equation}
Thus, the specific heat of the system is defined as
\begin{equation}\label{eq8}
C_{coop}=\frac{E_{coop}}{dT},
\end{equation}
where $T$ represents the temperature. We should mention here that DPT 
points separating the dynamically ordered and disordered phases are determined by benefiting
from the peaks of heat capacities. We also verified that the peak positions of heat capacities do not 
significantly alter when larger $L$ is selected.
\section{Results and Discussion}\label{discussion}

In this section, we will focus our attention on the nonequilibrium
dynamics of the mixed spin-1/2 and spin-3/2 Ising ferrimagnetic
system under a time dependent magnetic field. First of all,
we will discuss the dynamic nature of the system when
the system includes only the nearest-neighbor interaction between
spins, the single-ion anisotropy and external applied field.
Next, we will give and argue the global dynamic  phase diagrams including the
both dynamic critical and compensation temperatures in the case of the existence
the next-nearest neighbor interaction between spins-1/2 located on sublattice A.
Before we discuss the DPT features of the considered system,
we should notice that the situation of $h_{0}/|J_{1}|=0.0$ indicates the
equilibrium case, and our Monte Carlo simulation findings
for this value of $h_{0}/|J_{1}|$ are completely in accordance with the
recently published work \cite{Buendia2} where the equilibrium properties of the present system
were analyzed by following a numerical methodology of heat-bath Monte Carlo algorithm.

\begin{figure*}[!here]
\includegraphics[width=15.0cm,height=8.0cm]{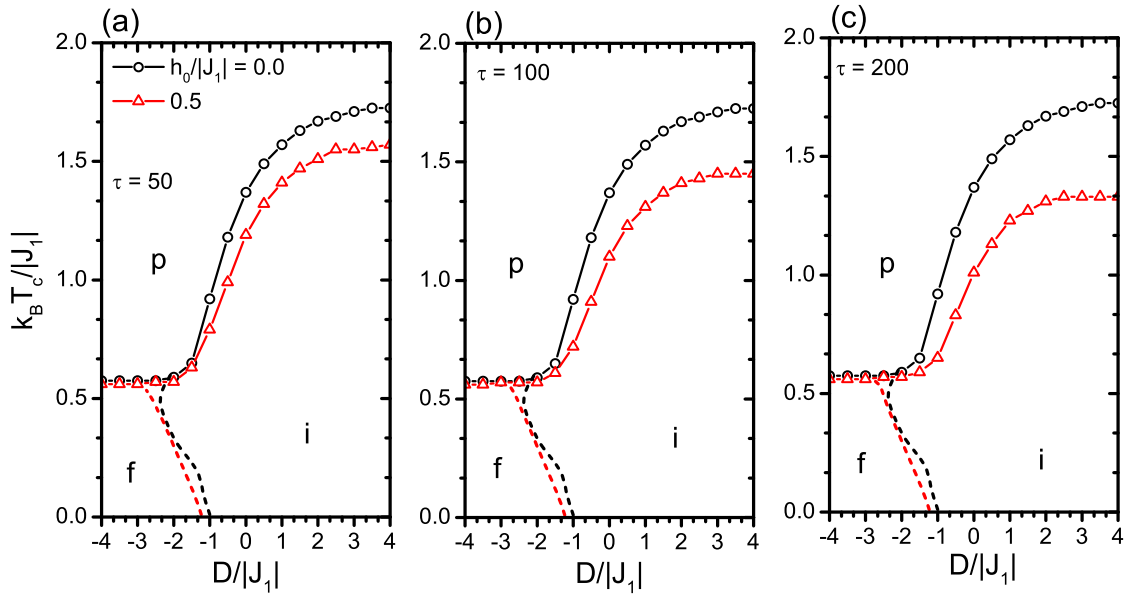}
\caption{(Color online) Dynamic phase boundaries of the system in the ($D/|J_{1}|-k_{B}T_{c}/|J_{1}|$)
plane with some selected values of external field amplitudes $h_{0}/|J_{1}|=0.0$ and $0.5$.
The curves are plotted for three values of oscillating period: (a) $\tau=50$, (b) $\tau=100$ and
(c) $\tau=200$. The dotted lines are boundary lines between two dynamically ordered phases.}\label{fig1}
\end{figure*}

The considered system exhibits  three types of magnetic behaviors depending on the Hamiltonian parameters.
These are dynamically ferrimagnetic (\emph{i}), ferromagnetic (\emph{f}) and paramagnetic (\emph{p}) phases,
respectively. In the first type of phase, namely in \emph{i} phase, $|Q_{A}|\neq|Q_{B}|$, and, the
time dependent sublattice magnetizations, $M_A(t)$ and $M_{B}(t)$ oscillate with
time around a non-zero value whereas they alternate around a non-zero value
and $|Q_{A}|=|Q_{B}|$ in the second type of phase, namely in  \emph{f} phase.
In \emph{p} phase which corresponds to the third type of phase,
$|Q_{A}|=|Q_{B}|$ and $M_A(t)$ and $M_{B}(t)$ oscillate around zero value, and they are
delayed with respect to the external applied magnetic field. Keeping in this mind,
we illustrate the dynamic phase diagrams in a $(D/|J_{1}|-k_{B}T_{c}/|J_{1}|)$ plane  with three
oscillation periods $\tau=50, 100$ and $200$ and for some selected
values of the  applied field amplitudes $(h_{0}/|J_{1}|)$ in Figs. \ref{fig1}(a)-(c).
One of the main findings is that DPT temperature decreases as the value of applied field amplitude increases.
The physical mechanism underlying this observation can be much better understood by following a simple way:
If one keeps the system in one well of a Landau type double well potential,
a certain amount of  energy coming from magnetic field is necessary to achieve
a dynamic symmetry breaking. If the amplitude of the applied field is less
than the required amount then the system oscillates in one well. In this situation,
the magnetization does not change its sign. In other words, the system oscillates around
a non-zero value corresponding to a dynamically ordered phase. As the temperature increases, the
height of the barrier between the two wells decreases. As a result of this, the less amount of
magnetic field is necessary to push the system from one well to another and hence
the magnetization can change its sign for this amount of magnetic field.
Consequently, the time averaged magnetization over a full cycle of the
oscillating magnetic field becomes zero. For the relatively high oscillation period
values, dynamic magnetizations corresponding to the instantaneous sublattice
order parameters can respond to the oscillating magnetic field with some delay, whereas
a competition occurs between the period $\tau$ of the field and the relaxation time of the
system as the period of the external magnetic field decreases. Hence, the dynamic magnetizations
can not respond to the external magnetic field due to the increasing phase lag between the field
and the time dependent magnetization. This mechanism makes the occurrence of the DPT
 difficult for the considered system. Another important observation is that an
unexpected sharp dip occurs  between dynamically ordered and disordered phases in the $(D/|J_{1}|-k_{B}T_{c}/|J_{1}|)$ plane
with increasing value of the amplitude of the external applied
field, and  our results show that observation such kind of  treatment explicitly depends upon
the value of $\tau$ of field. Furthermore,
it is necessary to state that for both large negative and positive values of
single-ion anisotropies, the phase transition points saturate a certain temperature regions,
and they tend to shift to the lower temperature regions with increasing amplitude and period of
the external applied field.

In Figs. 2(a)-(b), we depict the effect of the single-ion anisotropy on the thermal
variations of dynamic order parameters corresponding to the phase diagram illustrated in Fig. \ref{fig1}(a)
for value of $h_{0}/|J_{1}|=0.5$. It is clear from the figures that the treatments of the thermal variations
of sublattices as well as total magnetizations curves sensitively depend upon the value of single-ion anisotropy, for
selected values of Hamiltonian parameters. In the bulk ferrimagnetism of N\`{e}el, it is possible to classify the thermal
variation of the total magnetization curve in certain categories \cite{Neel}. According to this nomenclature, for $D/|J_{1}|\geq 0$,
the  considered system clearly points out a Q-type behavior, where the magnetizations of system begin to decrease
gradually starting from their saturation values with increasing thermal agitation, and then they vanish
at the DPT point.  One can easily see that, in the range $-1 < D/|J_{1}|< 0$, the
magnetizations tend to fall prominently from their saturation values, and the system
undergoes a second order  DPT as temperature increases. In addition to these, when $D/|J_{1}|<-1$, the system
exhibits a L-type behavior at which the total magnetization shows a temperature induced maximum which definitively
depends on the value of single-ion anisotropy as well as other Hamiltonian parameters.
Based on the above simulation observations, it is possible to make an inference that
the studied system has three types of dynamic magnetic behavior.
It is also worthy of note that even though the magnitudes of spins are different from each other, both
$Q_{A}$ and $Q_{B}$ exhibit a DPT at the same critical temperature, which is a result of
the nearest-neighbor exchange coupling $J_{1}$.

\begin{figure*}[!here]
\includegraphics[width=10.0cm,height=8.0cm]{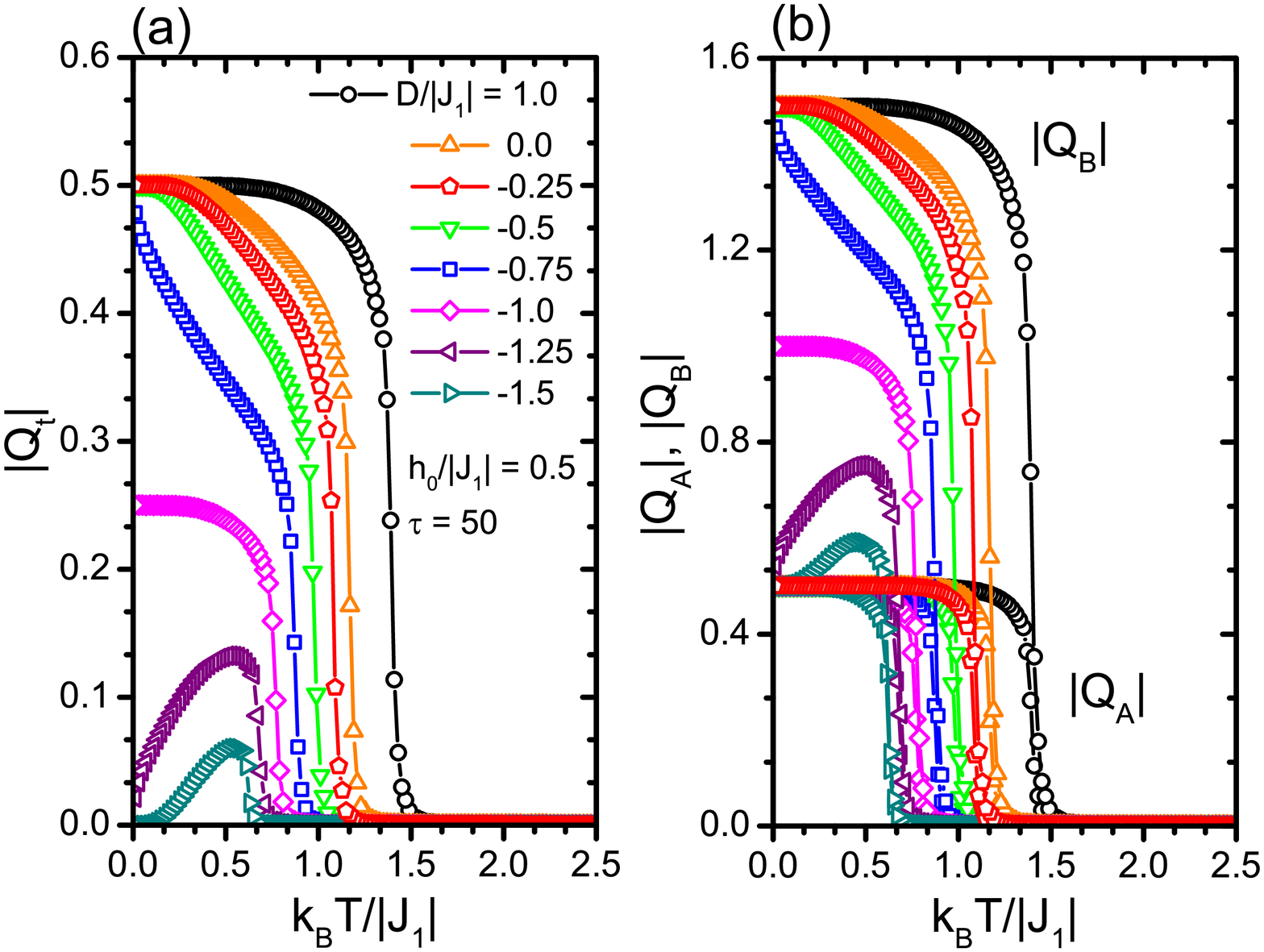}
\caption{(Color online) Effects of the single-ion anisotropy on the thermal variations of order parameters $|Q_{t}|, |Q_{A}|$ and $|Q_{B}|$
for a combination of Hamiltonian parameters corresponding to the phase diagram depicted in Fig. \ref{fig1}. }
\end{figure*}\label{fig2}

\begin{figure*}[!here]
\includegraphics[width=15.0cm,height=8.0cm]{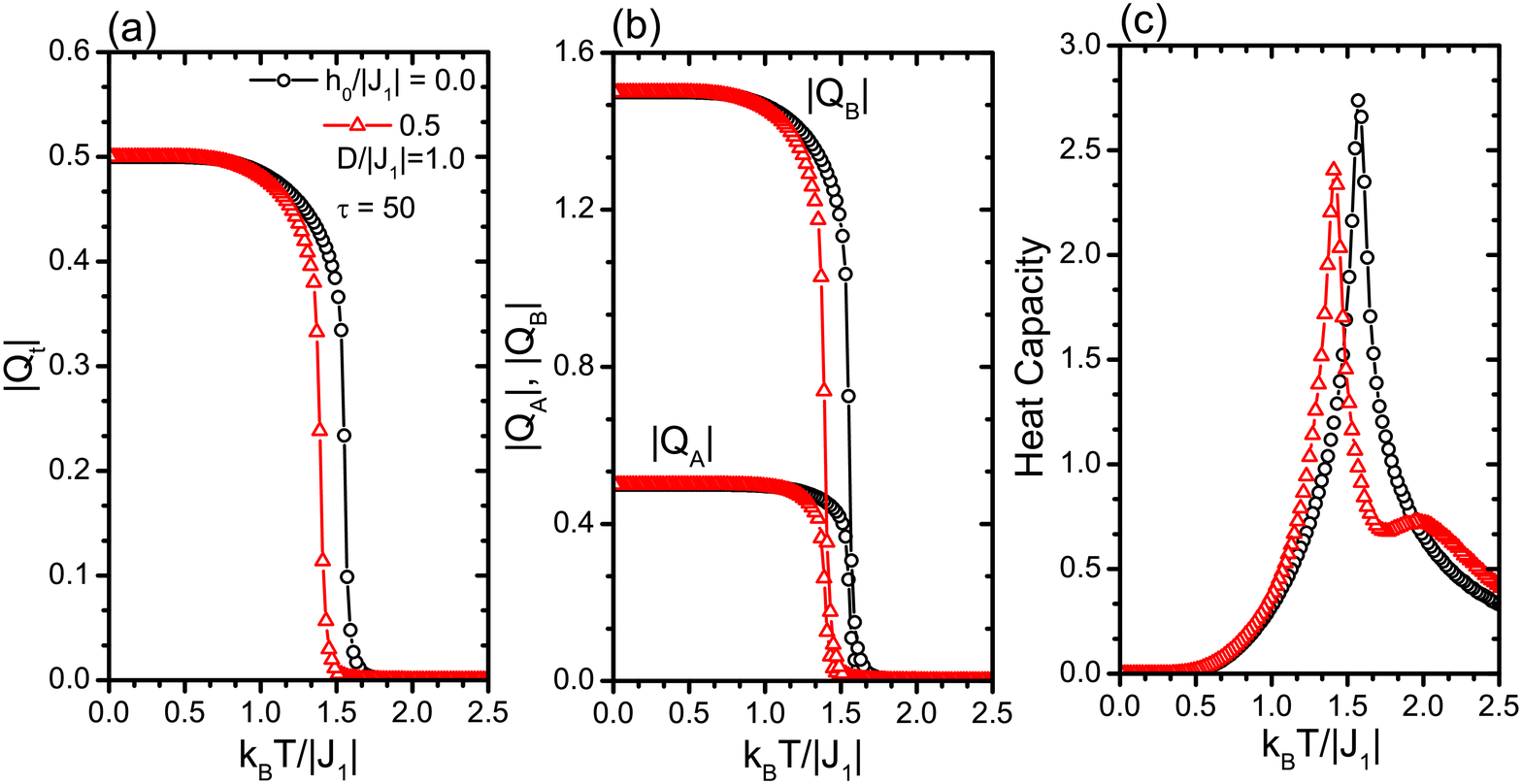}
\caption{(Color online) Temperature dependencies of (a) total
magnetization $|Q_{t}|$, (b) sublattice magnetizations   $|Q_{A}|$ and $|Q_{B}|$, and (c)
dynamic heat capacity for $D/|J_{1}|=1.0$ and $\tau=50$ with
$h_{0}/|J_{1}|=0.0$ and $0.5$.}\label{fig3}
\end{figure*}

The influences of the applied field amplitude $h_{0}/|J_{1}|$ on the
thermal variations of total and sublattice magnetizations as well as dynamic
heat capacity of the system are plotted in Figs. \ref{fig3}(a)-(c) corresponding to the phase
diagram constructed in Fig. \ref{fig1}(a) with value of single-ion anisotropy $D/|J_{1}|=1.0.$ In Fig.
\ref{fig3}(a), total magnetization curves of the system are shown. As seen in this figure, magnetization
curves exhibit Q-type behavior and dynamic critical temperatures decreases
with increasing $h_{0}/|J_{1}|$ values. On the other hand, dynamic heat capacity curves
which are depicted in Fig. \ref{fig3}(c) show a sharp peak behavior
indicating the phase transition temperature.  Moreover, one can readily see that increasing
value of the applied field amplitude gives rise to decrease  the maximum of
the dynamic heat capacity curves. We should note here that such kinds of dynamic
heat capacity behaviors have been found in a ferrimagnetic core-shell nanoparticle
composed of a spin-$3/2$ ferromagnetic core which is surrounded by a spin-1 ferromagnetic
shell layer under the presence of a time dependent magnetic field \cite{Yuksel, Vatansever}.

\begin{figure*}[!here]
\includegraphics[width=10.0cm,height=8.0cm]{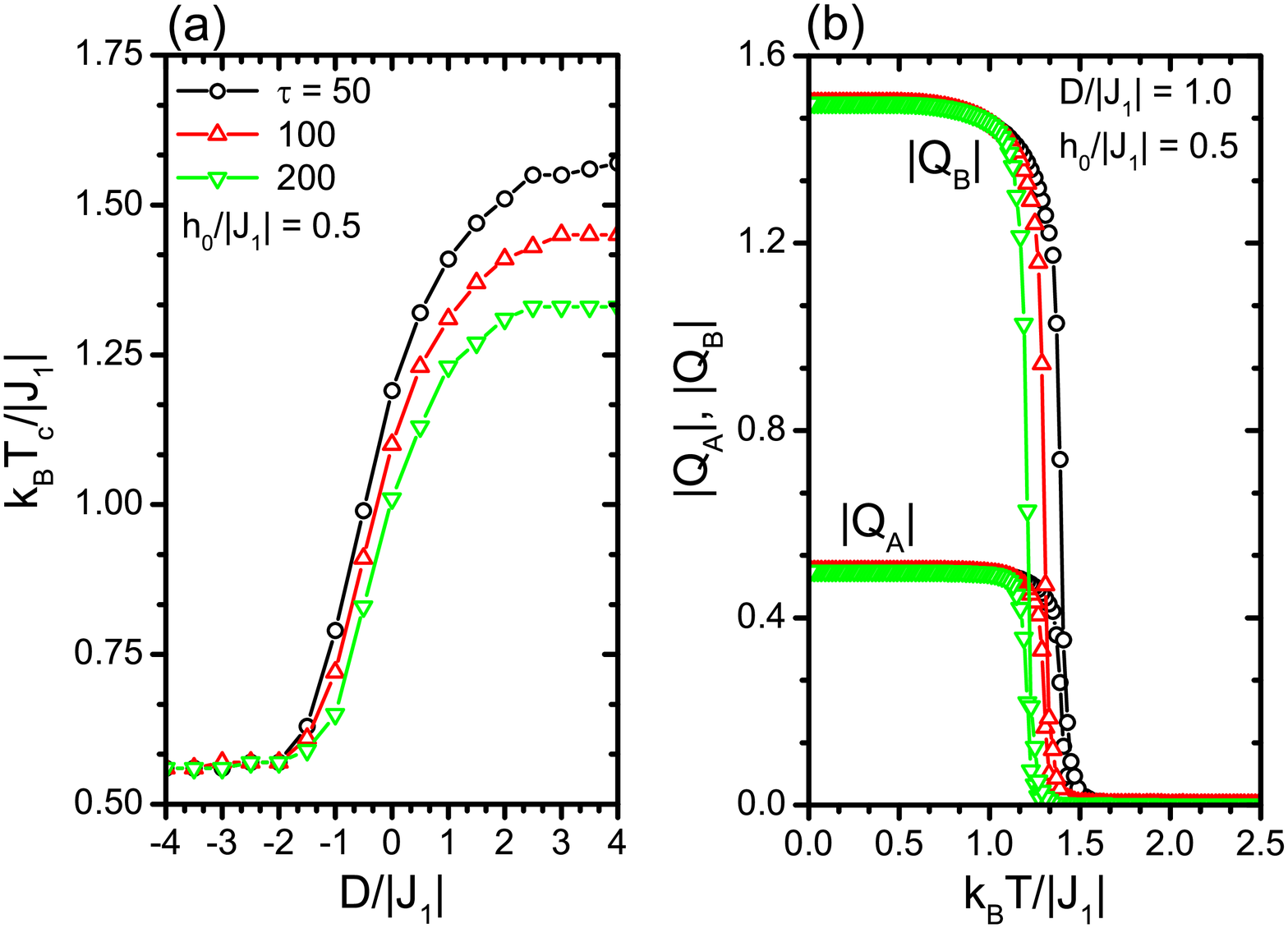}
\caption{(Color online) (a) Dynamic phase boundaries of the system
in ($D/|J_{1}|-k_{B}T_{c}/|J_{1}|$) plane for $h_{0}/|J_{1}|=0.5$ with $\tau=50, 100$ and $200$. (b) Effects of
the external applied field period on the thermal variations of sublattice magnetizations $|Q_{A}|$ and $|Q_{B}|$  for $D/|J_{1}|=1.0$
and $\tau=50, 100$ and $200$. }\label{fig4}
\end{figure*}

In Fig. \ref{fig4}, we investigate the effect of the applied field period on the DPT
features of the system. Phase diagrams in
Fig. \ref{fig4}(a) are constructed for a value of the applied field amplitude $h_{0}/|J_{1}|=0.5.$
It is possible to say that DPT
points are depressed with increasing applied field period especially in the
high values of the single-ion anisotropy.
The physics behind of these findings are identical to those emphasized
in Fig. \ref{fig1}. Therefore, we will not discuss these interpretations here.
Instead of this, in Fig. \ref{fig4}(b) we will give the influence of applied field period on
the thermal variations of sublattice magnetizations for a considered value
of single-ion anisotropy $D/|J_{1}|=1.0$ corresponding to the phase diagram
illustrated in Fig. \ref{fig4}(a). It is found that as the thermal
agitation increases starting from zero, the values of sublattice
magnetizations begin to gradually decrease and the system
undergoes a DPT at the critical temperature
which sensitively depends on value of the $\tau$.

According to our simulation results, the kinetic mixed spin-1/2 and spin-3/2 Ising
ferrimagnetic system including only nearest-neighbor interaction between spins,
the single-ion anisotropy and external applied field does not display a
dynamic multicritical behavior for a wide range of Hamiltonian parameters used in here,
in contrary to the previously published molecular field investigation where dynamic
first order phase transitions and dynamic tricritical points  have
been reported for the same model \cite{Deviren}.

\begin{figure*}[!here]
\includegraphics[width=15.0cm,height=8.0cm]{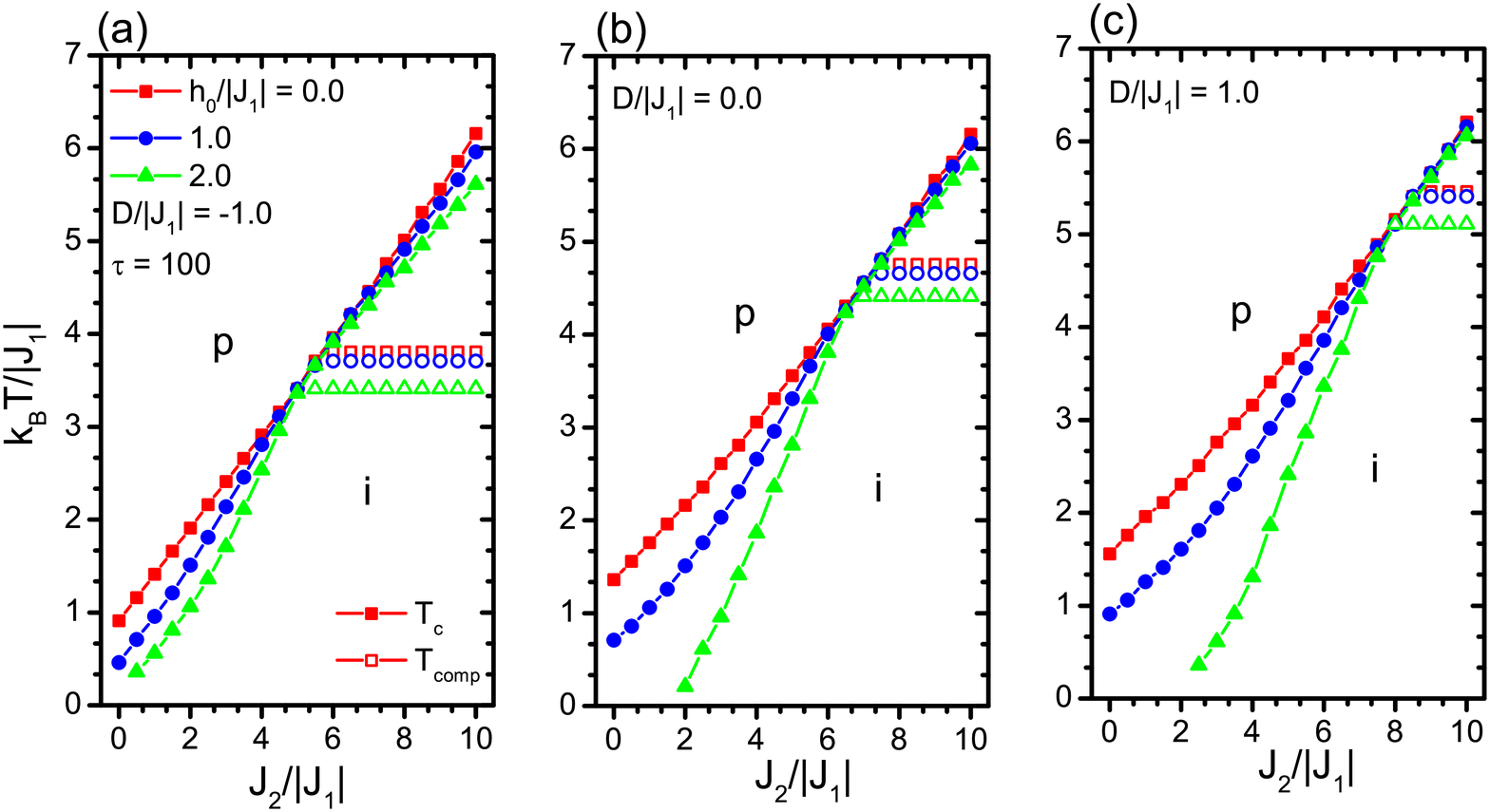}
\caption{(Color online) Dynamic phase boundaries including both critical and compensation temperatures
 in $(\mathrm{J_{2}/|J_{1}|-k_{B}T/|J_{1}|})$ plane for $\tau=100$ and
with some selected values of external field
amplitudes $\mathrm{h_{0}/|J_{1}|}=0.0, 1.0,$ and $2.0$. The curves are plotted for three values of single-ion
anisotropies (a) $D/|J_{1}|=-1.0$, (b) $D/|J_{1}|=0.0$ and (c) $D/|J_{1}|=1.0$. }\label{fig5}
\end{figure*}

\begin{figure*}[!here]
\includegraphics[width=15.0cm,height=8.0cm]{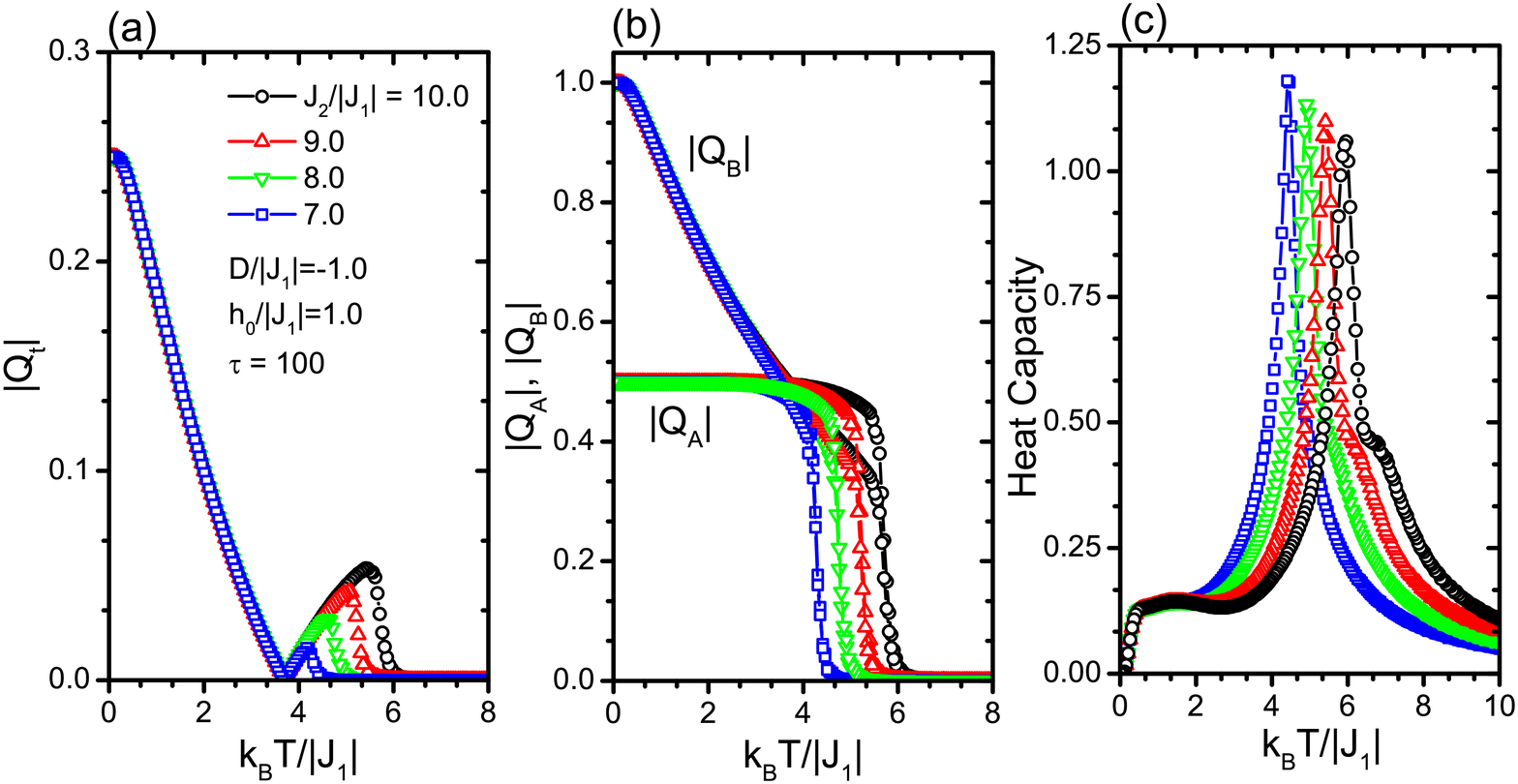}
\caption{(Color online) Influences of the next-nearest neighbor interactions on the thermal
variations of (a) total magnetization $|Q_{t}|$, (b) sublattice magnetizations $|Q_{A}|$ and $|Q_{B}|$ and (c)
dynamic heat capacity for some selected values of Hamiltoanian parameters corresponding to the phase
diagram illustrated in Fig. \ref{fig5}.  }\label{fig6}
\end{figure*}

In the following analysis, in order to  shed some light on the effect of the next-nearest neighbor interaction between
spins-1/2 of the studied system, we give the dynamic phase boundaries in $(J_{2}/|J_{1}|-k_{B}T/|J_{1}|)$
plane for some considered values of applied field amplitudes with $\tau=100$ in Figs. \ref{fig5}(a-c).
The phase boundaries containing both  dynamic critical and compensation temperatures
are plotted for three values of single-ion anisotropies $D/|J_{1}|=-1.0, 0.0$
and $1.0$, respectively. At first sight, one can clearly see that the dynamic compensation temperatures do not
emerge until $J_{2}/|J_{1}|$ reaches a certain amount of value. After the aforementioned value of $J_{2}/|J_{1}|$,
with an increment in $J_{2}/|J_{1}|$ does not lead to change the location of dynamic compensation
point  for fixed values of $D/|J_{1}|$ and $h_{0}/|J_{1}|$. In contrary to this behavior,
in accordance with the expectations, as the value of $J_{2}/|J_{1}|$ gets bigger starting from zero,
the much more thermal energy is necessary to reveal a DPT. On the other side, both
the dynamic critical and compensation temperatures strongly depend upon the selected values of $D/|J_{1}|$
and $h_{0}/|J_{1}|$. An increase in the value of $h_{0}/|J_{1}|$ gives rise to shift the
dynamic critical and compensation points to lower temperatures and also allows the system to display a
dynamic compensation behavior at the relatively low value of $J_{2}/|J_{1}|$. Additionally,
it is possible to make an inference that with increasing value of single-ion anisotropy,
the region where the dynamic compensation behavior occurs shifts to upward for
considered Hamiltonian parameters. This situation can be well understood
by comparing the Figs. \ref{fig5}(a), (b) and (c) with each other.

Effects of the next-nearest neighbor interactions on the
thermal variations of total and sublattice magnetizations as well as on dynamic heat capacity
of the studied system for some selected values of $D/|J_{1}|=-1.0, h_{0}/|J_{1}|=1.0$ with $\tau=100$
corresponding to the phase diagram shown in Fig. \ref{fig5}(a) are seen in Figs. \ref{fig6}(a-c).
We give the  total magnetization curves for changing value of $J_{2}/|J_{1}|$ in Fig. \ref{fig6}(a).
These curves explicitly refer the existence of a dynamic compensation behaviors, and they also exhibit a N-type
magnetic behavior. As we discussed before,  varying value of $J_{2}/|J_{1}|$ does not give rise to cause a
change in value of dynamic compensation point in temperature plane. Besides, in order to show how the dynamic
compensation phenomenon rises, thermal variations of the absolute values of
sublattice magnetizations are given in Fig. \ref{fig6}(b). It can be said that
an increase in value of $J_{2}/|J_{1}|$ leads to the existence of a dynamically
stronger ferromagnetic interaction between $\sigma$ spins. In this way,
the $\sigma$ spins can remain ordered at relatively higher
temperatures. When  the thermal energy increases starting from zero, the values of sublattice
magnetizations begin to gradually decrease from their saturation values
until both sublattices magnetizations are equail in
magnitude at a certain temperatures at which dynamic compensation point emerges below the
dynamic critical transition temperature. $J_{2}/|J_{1}|$ dependencies of dynamic heat
capacities are presented in Fig. \ref{fig6}(c). It is obvious from the figure that
the dynamic heat capacity curves exhibit a sharp peak at the transition temperature, and
when the value of $J_{2}/|J_{1}|$ increases, the dynamically ordered phase region gets wider, in
other words, the location of the sharp peak slides to a higher value in the
temperature plane. The shape of the heat capacity curves are also nearly same for
selected values of Hamiltonian parameters.

\begin{figure*}[!here]
\includegraphics[width=15.0cm,height=8.0cm]{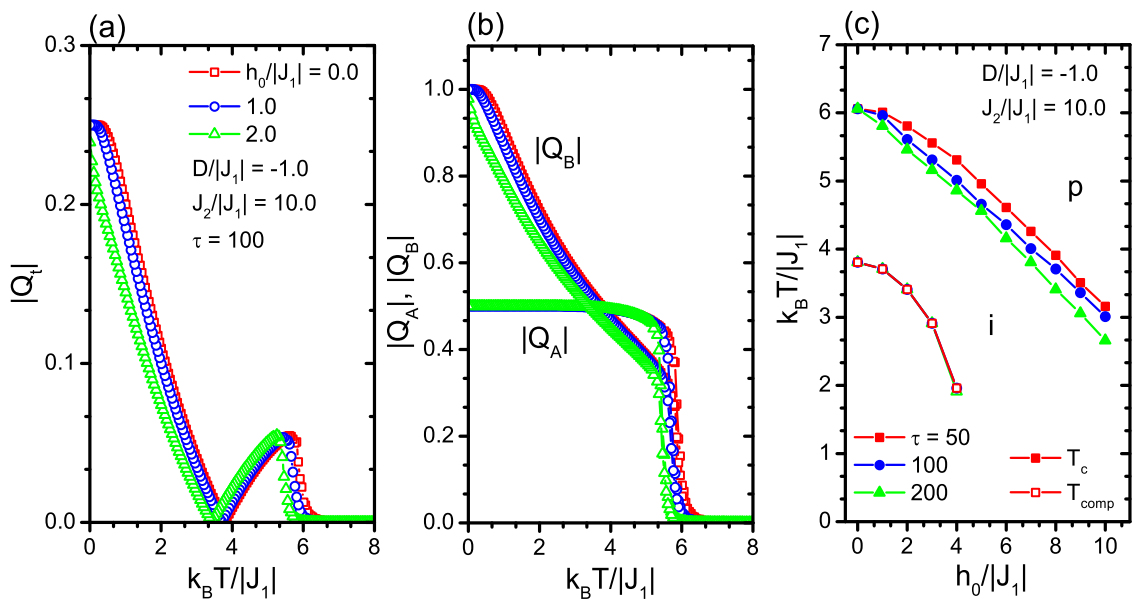}
\caption{(Color online) Effects of external applied field amplitude on the thermal variations of
order parameters $|Q_{t}|, |Q_{A}|$ and $|Q_{B}|$ for $D/|J_{1}|=-1.0, J_{2}/|J_{1}|=10.0$
and $\tau=100$ with $h_{0}/|J_{1}|=0.0, 1.0$ and $2.0$.}\label{fig7}
\end{figure*}

As a final investigation, we  give and discuss the influences of varying value of applied field
amplitude on the thermal variations of total and sublattices magnetizations exhibiting dynamic
compensation as well as critical temperatures in Figs. \ref{fig7}(a-b) corresponding to
the dynamic phase boundary seen in Fig.  \ref{fig5}(a) for $J_{2}/|J_{1}|=10.0.$ It is
possible to mention that both dynamic compensation and critical temperatures strongly depend on value of
the external applied field amplitude and they tend to shift to a lower region in temperature plane,
and compensation behavior disappears with increasing value of $h_{0}/|J_{1}|$. In order to demonstrate
the detailed magnetic behavior of system, we give the dynamic phase boundaries in ($h_{0}/|J_{1}|-k_{B}T/|J_{1}|)$ planes
for three values of the applied field periods such as $\tau=50, 100$ and $200$ for selected values of
single-ion anisotropy $D/|J_{1}|=-1.0$ and next-nearest neighbor interaction $J_{2}/|J_{1}|=10.0$
in Fig. \ref{fig7}(c). Based on the calculated phase diagrams,
it can be said that the decreasing (increasing) applied field period
has no effect on the dynamic behavior of compensation behavior
whereas it affects prominently the dynamic critical temperature of the system such that the dynamically
ordered phase region gets wider (narrower).

\section{Concluding Remarks}\label{conclusions}

In conclusion, it has been carried out a detailed Monte Carlo investigation based on standard
single-spin flip Metropolis algorithm to determine the true DPT
properties of a mixed spin-1/2 and spin-3/2 Ising ferrimagnetic system under a time varying
magnetic field. A complete picture of global dynamic phase diagrams separating the dynamically
disordered and ordered phases  has been constructed by benefiting
from the peaks of thermal variations of dynamic heat capacities in order to have a better understanding
of the physical background  underlying of the considered system. The most important observations reported
in the present study can be briefly summarized as follows:

\begin{itemize}
  \item When the considered system only includes the nearest-neighbor interaction, single-ion anisotropy and a
     time dependent sinusoidally oscillating magnetic field, it does not point out a dynamic compensation point.
  \item Stationary state solutions of the system strongly depend on the selected system parameters. As discussed in detail
  in previous section, with increasing values of amplitude and period of the external applied field, the dynamic phase boundaries
  tend to shift to the lower temperature regions in related planes, and a sharp dip occurs between dynamically
  ordered and disordered phases.
  \item In contrary to the previously published investigation for the same model where dynamic first-order phase
  transitions  and  tricritical points  have been reported\cite{Deviren},
  it has not been found any evidence of the dynamic first-order phase transitions in our present work.
  The reason is most likely the fact that the method we used completely takes into account the thermal
  fluctuations, which allows us to obtain non-artificial results.
  \item When the next-nearest neighbor interaction between spins-1/2 is included and exceeded a
  characteristic value which sensitively depends on value of the single-ion anisotropy and amplitude of
  the external applied field, the system exhibits a dynamic compensation behavior below its critical
  temperature. According to the our simulation results, the changing  value of applied field period has
  no effect on the location of dynamic compensation point.
\end{itemize}

Finally, we should note that it is possible to improve the obtained results by making use of a
more realistic system such as Heisenberg type of Hamiltonian. From the theoretical point of view,
such an interesting problem may be subject of a future work in order to provide deeper understanding of
ferrimagnetic materials under  a time dependent alternating magnetic field source.

\section*{Acknowledgements}
The numerical calculations reported in this paper were
performed at T\"{U}B\.{I}TAK ULAKBIM (Turkish agency),
High Performance and Grid Computing Center (TRUBA
Resources).

\end{document}